# Room temperature polariton condensation from Whispering gallery modes in CsPbBr$_3$ microplatelets

*Laura Polimeno*[1†], *Annalisa Coriolano*[1†], *Rosanna Mastria*[1*], *Francesco Todisco*[1], *Milena De Giorgi*[1], *Antonio Fieramosca*[1], *Marco Pugliese*[1], *Carmela T. Prontera*[1], *Aurora Rizzo*[1], *Luisa De Marco*[1], *Dario Ballarini*[1], *Giuseppe Gigli*[1,2], *Daniele Sanvitto*[1*]

[1]CNR Nanotec, Institute of Nanotechnology, via Monteroni, 73100, Lecce, Italy.
[2]Dipartimento di Matematica e Fisica "Ennio de Giorgi", Universitá del Salento

## Abstract

Room temperature (RT) polariton condensate holds exceptional promise for revolutionizing various fields of science and technology, encompassing optoelectronics devices to quantum information processing. Using perovskite materials like all-inorganic CsPbBr$_3$ single crystal provides additional advantages, such as ease of synthesis, cost-effectiveness, and compatibility with existing semiconductor technologies.

In this work, we show the formation of whispering gallery modes (WGM) in CsPbBr$_3$ single crystals with controlled geometry, synthesized using a low-cost and efficient capillary bridge method. Through the implementation of microplatelets geometry, we achieve enhanced optical properties and perfor-mance thanks to the presence of sharp edges and a uniform surface, effectively avoiding non-radiative scattering losses caused by defects. This allows us not only to observe strong light matter coupling and formation of whispering gallery polaritons, but also to demonstrate the onset of polariton condensation at RT. This investigation not only contributes to the advancement of our knowledge concerning the exceptional optical properties of perovskite-based polariton systems, but also unveils prospects for the exploration of WGM polariton condensation within the framework of a 3D perovskite-based platform, working at RT.

---

[†] These authors contributed equally: L. Polimeno, A. Coriolano
[*] e-mail: rosanna.mastria@nanotec.cnr.it, daniele.sanvitto@nanotec.cnr.it

The unique characteristics of polariton condensate, including low excitation thresholds and ultrafast dynamics, open up unique opportunities for advancements in photonics and optoelectronics devices.

## Introduction

The impact of all-optical technology is significant in various areas[1], including communication[2], data storage[3], renewable energy[4], and security[5]. Exciton-polaritons are bosonic quasi-particles arising from the strong coupling between excitons and photons, inheriting unique properties from their bare components[6]. These properties can be harnessed for the realization of compact and integrated devices with wide tunability[7].

One of the notable advantages of polariton technology is the occurrence of polariton condensate at significantly lower excitation thresholds compared to conventional laser devices. Consequently, polariton lasers can operate at lower energy input levels, resulting in enhanced energy efficiency and potentially reducing power consumption in optoelectronic devices[8,9,10]. The critical factor for realizing device integration and commercialization lies in the operation of polariton condensate systems at RT. This can be achieved by exploiting the strong binding energy of different type of active materials in polaritons platform, such as organic materials[11], transition metal dichalcogenides[12,13], and perovskites[14,15,16].

Cesium Lead Bromide ($CsPbBr_3$) stands out as a highly promising semiconductor material for polariton-based devices, finding applications in optoelectronics[17,18], photovoltaics[19], and light-emitting devices[20,21]. $CsPbBr_3$ exhibits a perovskite crystal structure with general formula $ABX_3$ in which cesium (Cs) cations occupy the larger A-site, lead (Pb) cations occupy the B-site, and bromine (Br) anions coordinate the Pb cations to form an octahedral arrangement[22]. $CsPbBr_3$ has a direct bandgap and its strong exciton binding energy allows the formation of stable excitons even at RT, enabling various excitonic phenomena such as polariton formation and low threshold condensate[23,24,25].

As a single crystal, $CsPbBr_3$ demonstrates remarkable optical and electronic properties, making it an attractive material to design customized microcavity geometries, exploiting different type of confined modes, such as whispering gallery and Fabry-Perot (F-P) cavity modes[26]. Depending on the geometry of the crystals, perovskites can effectively serve as intrinsic high-quality microcavities without additional device implementation[14]. For instance, the four smooth lateral faces of square crystals or the two parallel lateral faces of microwires can act as mirrors to support WGM[23,27,28] or F-P lasing[29,30,31], respectively[32]. In this context, WGM microcavities emerges as a promising photonic platform possessing high quality (Q) factors and ultra small size. These enable enhanced light-matter interactions and offer superior optical confinement, facilitating low threshold lasing[23]. However, despite the promising initial reports and the attractive features of intrinsic WGM microcavities in perovskite crystals[23,33,34], RT exciton polaritons and polariton condensation have not yet been demonstrated in this photonic platform.

In this framework, one-dimensional multimode WGM exciton-polariton condensation has been demonstrated in zinc oxide microwires at cryogenic temperature (T = 10 K)[35], while RT WGM polariton condensation is achieved in a GaN hexagonal wire[36]. The constraint within these current studies stems from the challenge of effortlessly controlling and engineering the geometrical shape of the active material during the fabrication process and the scalability of the process, particularly for seamless integration into practical devices. In this regard, the ease of handling perovskites makes them a promising candidate for such applications.

Nevertheless, achieving high-quality WGM in $CsPbBr_3$ single crystals requires precise control over the crystallization process to ensures well-defined geometric features and smooth faces, allowing photons to circulate and resonate efficiently via self-interference. However, strategies for shape engineering targeting $CsPbBr_3$ crystals with defined geometry for the tuning of the WGM microcavity effect, remain limited. Given the challenges arising from the differing solubility of CsBr and $PbBr_2$ precursors, CVD has been extensively investigated for growing $CsPbBr_3$ microplatelets.[37] However, the need for high temperatures, the demand for specific substrates to ensure a proper lattice match, and the limited control over the crystal's shape, make CVD inadequate for optimizing WGM $CsPbBr_3$ crystals. In this work, we employ a low-cost and versatile synthesis method based on the capillary bridge approach[38,39] to grow $CsPbBr_3$ square single crystals with smooth surface and lateral facets to control the resulting WGM properties, obtaining compelling evidence of polariton condensation at RT. Exploiting the high optical quality of as synthesized $CsPbBr_3$ single crystal microplatelets, we unambiguously demonstrate the onset of RT whispering gallery polariton condensation. Compared to the previously reported lasing thresholds in the $CsPbBr_3$ structure[40,41,42], our low polariton condensation threshold ($\simeq$ 8 $\mu J/cm^2$) is achieved through the formation of polariton WGM in the single crystal microplatelets.

Our experimental observations reveal the emergence of three distinct coherent states at different power thresholds, emitted from the crystal edges.

Furthermore, through manipulation of the CsPbBr$_3$ geometry we measure the exciton polariton dispersion from a one-dimensional grating directly written on the top of the crystal.

These experimental results provide nuanced insights into the debated kinetic processes governing perovskite-based systems. Consequently, they reveal novel prospects for probing and modulating macroscopic quantum coherent states within 3D perovskite-based polariton systems by controlling the geometrical shape of CsPbBr$_3$ single crystals. The contemplation of such phenomena stands to significantly advance the fundamental understanding and practical utilization of polariton technologies into uncharted scientific frontiers.

## Result and Discussion

CsPbBr$_3$ single crystal microplatelets with square shape are successfully grown using an improved capillary bridge approach which combines the nanoconfinement of the precursors solution with the antisolvent assisted crystallization (Fig. 1a).

The ability to control nucleation and growth in CsPbBr$_3$ remains a quite challenging task especially for reaching controlled geometries. Theoretically, CsPbBr$_3$ crystals should grow as a perfect 3D cube, with all six {100} crystal faces growing at an identical rate, aiming for an equilibrium morphology with the smallest surface energy. However, in practice, several factors like the impurities on the substrate and/or variation of the environmental conditions (i.e. temperature and humidity) can influence the growth rates of different facets[37,43].

To achieve an ideal 2D square morphology (i.e. microplatelets), the capillary bridge method stands out as one of the most promising among solution-based approaches. This method has been proven highly effective for hybrid metal halide perovskites[44,45], but, up to now, the growth of CsPbBr$_3$ square microplatelets using this approach has not been demonstrated. We use a silicon template structured with periodic square pillars and featuring asymmetric wettability (see Fig. S1 in Section I of Supplementary Materials). This ensures successful nanoconfinement of the precursor liquid at the positions of the micropillars and at the same time restricts the growth of two of the six facets of the crystal, promoting the development of microplatelets with controlled thickness (see Section I of Supplementary Materials). However, the confinement of the precursor solution does not guarantee controlled crystallization of single-crystals or the controlled growth of the remaining four crystal faces. Therefore, we manipulate the nucleation, and consequently, the crystal growth by directly acting on the solution environment. The CsPbBr$_3$ precursor solution is typically prepared using CsBr and PbBr$_2$ in a stoichiometric ratio, with dimethylsulfoxide (DMSO) as the solvent. The DMSO ensures effective and uniform dissolution of the precursors, facilitating the precipitation of high-quality crystals[46,47]. However, because of the strong interaction between DMSO and the precursors[48,49], other strategies, such as temperature adjustments or variation of the concentration, are required to achieve saturation of the monomer and to initiate nucleation. To this aim, we introduce an antisolvent into the precursor solution. We find that the use of an antisolvent, especially chlorobenzene, facilitate control over the saturation of the precursor solution and the crystallization kinetics, enabling the growth of high-quality crystals capable of acting as WGM microcavities. The resulting microplatelets, having lateral dimension of ∼ 6 $\mu$m, display a regular shape, smooth surface, and distinct edges, as evident from the optical micrographs images in both bright and dark fields, as well as from the scanning electron microscope (SEM) image in Fig. 1b. A comprehensive description of the pa-rameters influencing the growth of CsPbBr$_3$ microplatelets can be found in Section I of Supplementary Materials. After achieving good-quality microplatelets, an optical study are conducted to investigate the optical confinement of the as-grown crystals. We study the nonlinear response of the system under pulsed laser excitation (with a 145-fs pulsed laser at 2.64) and a spot size covering the entire microplatelet (see Methods for further information).

Fig. 1c-f show the power dependence of the emission from a 6x6 µm$^2$ square microplatelet in both real and energy space. The incident fluence is evaluated as: $F = \frac{P \cdot abs}{RR \cdot (\pi r^2)}$, where P is the incident power, abs is the absorption coefficient at the energy of the laser (2.64 eV - 470 nm) extracted from the CsPbBr3 absorption (Fig. S5 in the Supporting Material), RR is the repetition rate of the pulsed laser (10 kHz) and r is the laser spot radius.

At low fluence (F = 1.3 µJ/cm$^2$) shown in Fig. 1c, the emission is uniform and homogeneous throughout the entire structure, featuring a single peak centered at 2330 meV with a full width at half maximum (FWHM) of ∼ 70 meV. As the incident pump power is increased, we observe a first threshold at F ≃ 8 µJ/cm$^2$, where an intense emission peak appears at about E = 2317 meV. At this point, the real-space emission primarily originates from the crystal edges. We attribute this peak to the onset of the polariton condensation, as indicated by the narrowing of the emission intensity linewidth (FWHM ≃ 4 meV). By continuously increasing the incident fluence, two additional peaks appear at 2304 meV (F ≃ 9.5 µJ/cm$^2$) initially, and then at 2290 meV (F ≃ 16 µJ/cm$^2$). At high fluences (Fig. 1e and 1f), the real-space emission becomes completely concentrated at the square cavity edges, and the central part of the square is entirely dark.

The evenly spaced peaks in the energy spectrum demonstrate the formation of the polariton condensate from WGM, supported by the 150 nm-thick perovskite microplatelet[23](see Section III of Supporting Materials).

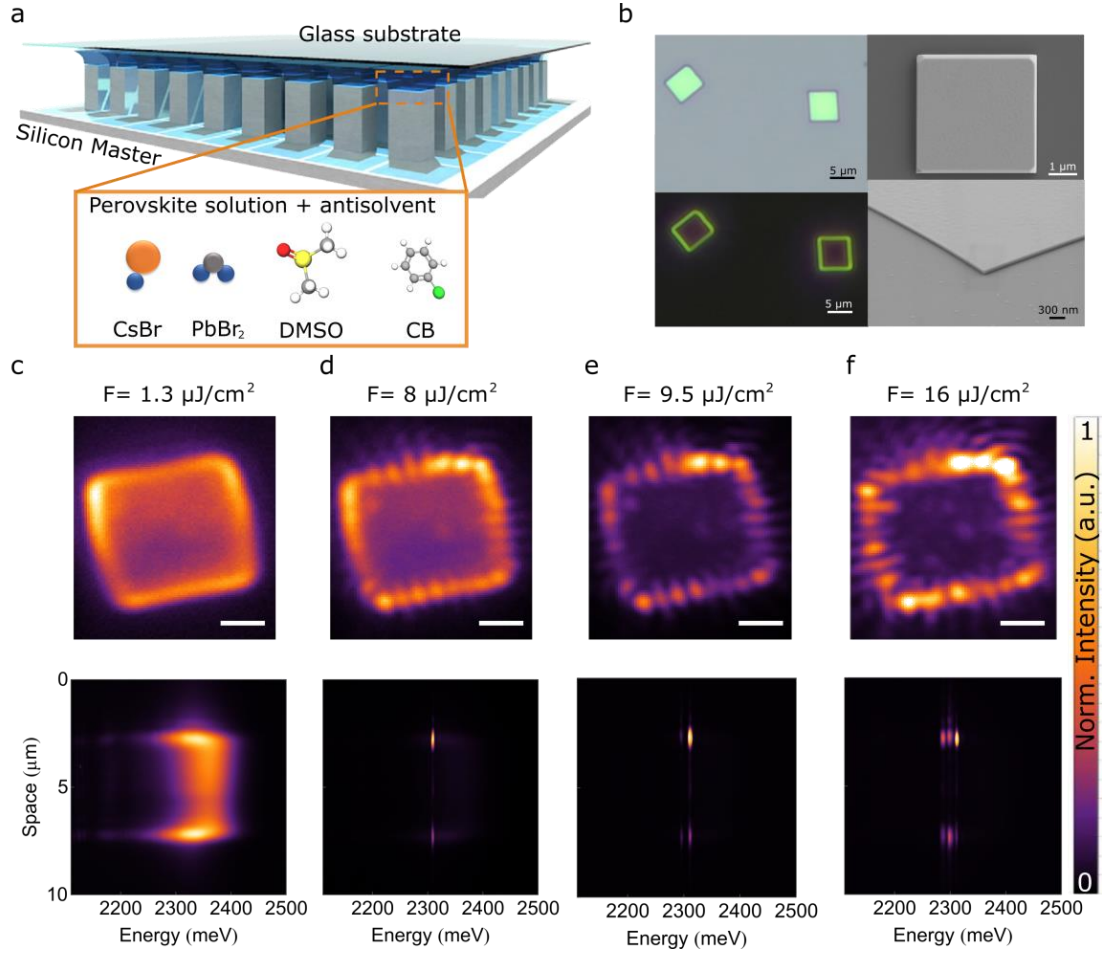

Figure 1: (a) Sketch of the synthetic method. (b) Optical micrograph and SEM images of CsPbBr$_3$ microplatelets. (c-d-e-f) Real space map and corresponding photoluminescence spectra for four different incident pump fluences, exciting a 150 nm-thick perovskite microplatelet (145-fs pulsed excitation at 2.64 eV). At low fluence, the emission from the microplatelet appears uniform and homogeneous in the real space map, exhibiting a single peak centered at 2330 meV, while the edge photoluminescence shows a redshift. Above the condensate threshold, the edge emission becomes significantly stronger than the emission from the center area, and three distinct narrow polariton peaks emerge at different thresholds.

To further corroborate the evidence of polariton condensation in the three polariton modes (Fig. 3a), we directly analyze the integrated emission intensity as a function of pumping fluences, as shown in Fig. 3b. The nonlinear increase in emission intensity is clearly
8

evident for all three modes, with distinct sharp rises at three thresholds indicated by the red (F ≃ 8 $\mu$J/cm$^2$ for mode 1), black (F ≃ 9.5 $\mu$J/cm$^2$ for mode 2), and blue dashed lines (F ≃ 16 $\mu$J/cm$^2$ for mode 3), respectively.

Moreover, the observed blueshift in the polariton condensation (Fig. 3c) is a unique phenomenon associated with the strong coupling between excitons and the confined optical modes in the perovskite crystal which exclude the possibility of observing photonic lasing. Specifically, as the incident fluence increases, more excitons and photons are generated, leading to a higher density of polaritons in the system. This increased density results in a higher probability of polariton-polariton interactions, and these interactions contribute to the blueshift observed in the polariton condensation emission. The blueshift in the emission spectrum provides valuable insights into the behavior of polaritons under strong light-matter coupling conditions and is a significant characteristic of polariton condensates in such systems.

To further validate this interpretation, time-resolved photoluminescence (TRPL) measurements are conducted below and above the polariton condensation threshold (Fig.3), in order to compare the different kinetic dynamics of the system. At low excitation power (F ≃ 1.3 $\mu$J/cm$^2$), the lifetime of the photoluminescence is analyzed over time by sending the detected signal into a streak camera. The time decay is fitted with a bi-exponential decay, revealing a long lifetime of $t_1$ ∼ 250 ps, which is consistent with previous values reported in the literature[50].

At higher fluences (F ≃ 30 $\mu$J/cm$^2$), the three polariton modes emit at different energies, enabling the discrimination of their three decay times. The decay time of the three peaks at higher energy is faster compared to that of the exciton (∼ 250 ps), falling below the resolution of the streak camera (∼ 2 ps). The ultrafast decay of the three modes further supports the occurrence of polariton condensation at different threshold.

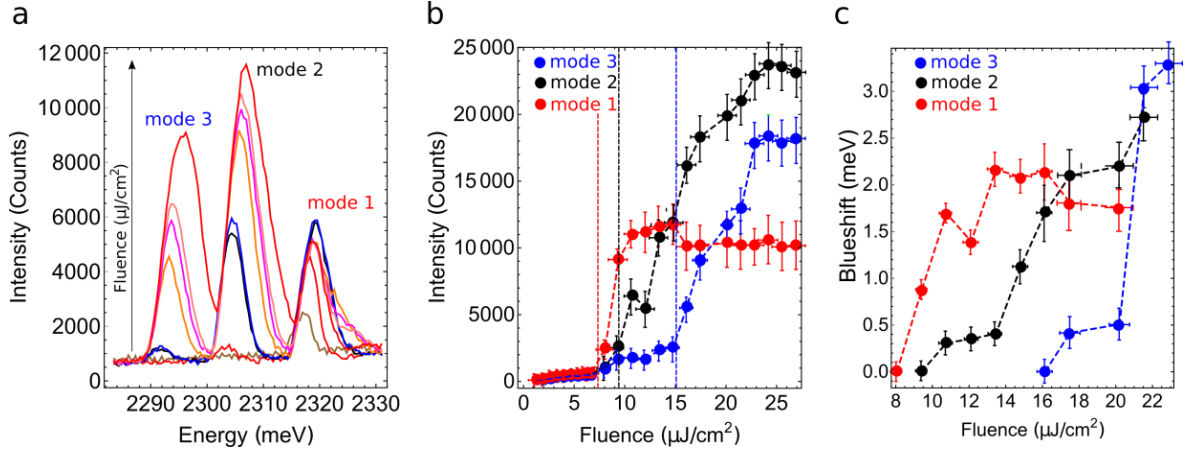

Figure 2: (a) Comparison of the emission spectra with increasing incident fluences, revealing the formation of the polariton condensate for three distinct polariton modes with three different thresholds. (b) Integrated intensity of the emission as function of incident pump fluences for mode 1 (red dots), mode 2 (black dots) and mode 3 (blue dots). Three different thresholds are clearly visible: the first one at F ≃ 8 $\mu$J/cm$^2$ (red dashed line) indicates the polariton condensate of the higher energy mode (mode 1), the second one at F ≃ 9.5 $\mu$J/cm$^2$ (black dashed line) marks the onset of the polariton condensate of mode 2, while the mode 3 condenses at F ≃ 16 $\mu$J/cm$^2$.

Finally, in order to directly observe the polariton modes beyond the light line, we synthesized a platelet with one side at least 10 times longer than the one reported previously. This allows to directly write a 1D- TiO$_2$ grating on top of the CsPbBr$_3$ single crystal. To this extent we follow a slightly modified capillary bridge approach reported recently[51] (see Section IV of Supporting Materials for further details).

Fig. 4a displays a scanning electron microscope (SEM) image of the grating fabricated on top of the perovskite using a low-dose electron beam lithography (see Methods for fabrication details). The periodic grating is designed to allow the extraction of the signal beyond the air light-line, specifically the signal arising from total internal reflection (TIR) at the crystal-air interface.

The extended platelet is off-resonantly pumped with a continuous-wave laser with a photon energy of 2540 meV. The photoluminescence energy dispersion (Fig. 4b) shows the lower polariton branches of a 400 nm-thick single crystal of $CsPbBr_3$, distinguishing between the Transverse Electric (TE) and Transverse Magnetic (TM) components. Fig. 4b exhibits the typical dispersion of the exciton-polariton, with the optical resonances bending close to the excitonic resonance at large in-plane wavevector[14]. Particularly, the presence of propagating and counter-propagating lower polariton branches results from the strong coupling of the lowest-energy excitons (E$_{exc}$ = 2408 meV) with different optical modes, as determined by the microcavity free spectral range. In Figure 4b, the blue dashed lines represent the theoretical lower polariton branches resulting from the strong coupling between the exciton transition (red dashed line) and the TE and TM polarized cavity modes, with Rabi splittings of $\Omega_{TE}$ = 250 meV and $\Omega_{TM}$ = 140 meV for TE and TM modes, respectively (see Methods for further information).

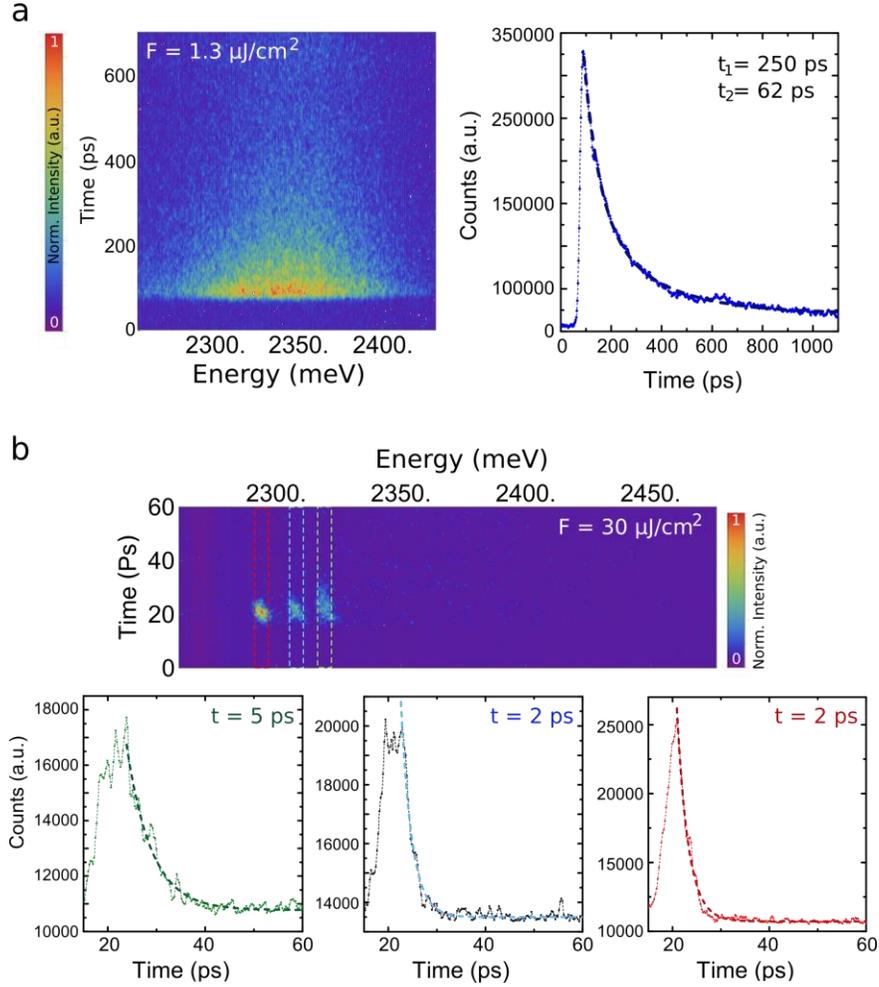

Figure 3: (a) Below the threshold, the TRPL decay depends on the exciton lifetime. The decay is fitted with a bi-exponential model, yielding a long time component of 250 ps and a shorter one of 62 ps. (b) Above the threshold, all three peaks exhibit a short lifetime, surpassing the resolution limit of ∼ 2 ps, confirming the onset of the polariton condensate for the three polariton modes.

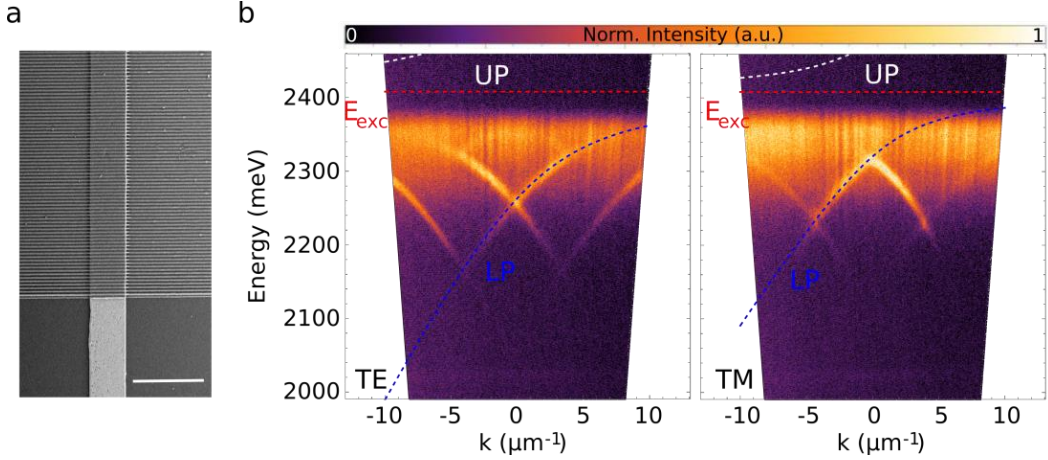

Figure 4: (a) Directly-written TiO$_2$ grating on a 400 nm-thick CsPbBr$_3$ wire, captured in a SEM image. Scale bar: 10 µm (b) Energy Vs in-plane momentum photoluminescence of the wire shown in panel (a), in which TE and TM lower polariton modes are clearly visible. The blue dashed lines are the theoretical dispersion of the eigenstates of the coupled system resulting from the strong coupling between the excitonic transition at E$_{exc}$ = 2408 meV (red dashed line) and the optically confined modes within the wire. The upper polariton modes (white dashed lines) are not experimentally measurable.

## Conclusion

In conclusion, our work demonstrates the formation of polariton condensation at RT in CsPbBr$_3$ single crystals with controlled geometry, achieved through the optimization of the asymmetrical-wettability synthesis method. Notably, the implementation of mi-croplatelets geometry further enhances the optical properties and performance of the perovskite material.

The power dependence of emission from the microplatelets reveals distinct polariton condensates at different thresholds. The blueshift observed above threshold is a unique phenomenon resulting from strong coupling between excitons and confined optical modes in the perovskite crystal. Time-resolved photoluminescence measurements provide further confirmation, showcasing the ultrafast decay of the three polariton modes, solidifying the onset of the polariton condensation at different thresholds.

Our study delves into the dynamic behaviors of perovskite-based systems, revealing rich insights. Furthermore, it unveils promising avenues for investigating macroscopic quantum coherent states originating from WGM in $CsPbBr_3$ perovskite polariton systems at RT. The exceptional properties of polariton condensation, including low excitation thresholds and ultrafast dynamics, hold substantial promise for advancing optoelectronics, high-speed optical communications, and RT technological applications.

## Methods

### Sample Fabrication

Dimethyl sulfoxide (DMSO), Acetone, Chlorobenzene (CB), Dichloromethane (DCM) and heptadecafluorodecyltrimethoxysilane (FAS) are purchased from Sigma Aldrich. Lead (II) Bromide ($PbI_2$) and Cesium bromide (CsBr) are purchased from Alfa Aesar. All chemicals are used as received without any further purification.

Microstructured silicon templates with asymmetric wettability are employed to control the assembly of $CsPbBr_3$ single crystals. The $CsPbBr_3$ solution (0.02 M) is prepared by directly dissolving stoichiometric CsBr and $PbBr_2$ in a DMSO/CB mixture (1:0.4). A 1 $\mu$l aliquot of the precursor solution is dispensed onto the selectively functionalized silicon master and then covered with a glass substrate, which has been previously cleaned with acetone and IPA in an ultrasonic bath and treated with oxygen plasma for 10 minutes. Using a home-built apparatus, uniform pressure is applied to the top of the sandwiched substrates. The growth initiated in an oven with controlled humidity (RH 30%) at 30°C for 48 hours, followed by 80°C for 12 hours.

The microwire single crystal is spin-coated with Polymethyl methacrylate (PMMA) 950k at high accelerations and soft-bake. Then, the sample is immediately spin coated with a thin surfactant-enhanced discharge layer and placed in the Electron Beam Lithography (EBL) vacuum chamber without any other baking step. The pattern is exposed with a 30kV electron beam, using a positive tone process at low doses; the sample is finally developed with a low-stress/low-dose developer, followed by a surfactant-enhanced deionized water rinsing and $N_2$ drying.

A 60 nm-$TiO_2$ film is deposited via electron beam evaporation on top of sample, in absence of any oxigen flow and keeping the sample at RT. The $TiO_2$ grating is obtained after the lift-off in chlorobenzene of the residual PMMA.

## Optical Measurements

All the optical measurements reported in the work are performed under ambient conditions at RT.

For the photoluminescence measurements, the microwires and microplatelets are off-resonant excited by using a continuous-wave 2540 meV diode laser. The photoluminescence is recorded in reflection configuration, using a 100x objective with NA=0.9.

For nonlinear measurements, a tunable femtosecond laser (with pulse width ∼ 145 fs, repetition rate 10 kHz) is used to excite the microplatelets. The energy of the laser and the corresponding real space spot size dimensions are ∼ 2.64 eV and ∼ 10 $\mu m^2$, respectively.

## Simulations and fits

The experimental data reported in Fig. 4 of the main text are fitted with a 2X2 coupled harmonic oscillator Hamiltonian:

$$H_k = \begin{pmatrix} E_{ph}(k) & \Omega/2 \\ \Omega/2 & E_{exc} \end{pmatrix}$$

where $E_{exc}$ is the exciton energy, $\Omega$ is the Rabi splitting and $E_{ph}(k)$ is the energy dispersion of the photonic branch.

## Acknowledgements

The authors gratefully thank P. Cazzato and S. Carallo for technical support.

This work was supported by the Italian Ministry of University (MUR) for funding through the project 'Hardware implementation of a polariton neural network for neuromorphic computing'—Joint Bilateral Agreement CNR–RFBR (Russian Foundation for Basic Research)–Triennal Program 2021–2023, the MAECI project 'Novel photonic platform for neuromorphic computing', Joint Bilateral Project Italia–Polonia 2022–2023, PNRR MUR project PE0000023-NQSTI, FINANCED BY THE EUROPEAN UNION – Next Generation EU, PNRR MUR project IR0000016-I-PHOQS, FINANCED BY THE EUROPEAN UNION – Next Generation EU, project 'Progetto Tecnopolo per la Medicina di precisione', Tecnomed 2 (grant no. Deliberazione della Giunta Regionale n. 2117 del 21/11/2018), the MUR project "ECOTEC - ECO-sustainable and intelligent fibers and fabrics for TEChnic clothing", PON « RI» 2014–2020, project N° ARS0100951, CUP B66C18000300005.

# Supplementary Materials: Room Temperature polariton condensation from Whispering gallery modes in CsPbBr$_3$ microplatelets

*Laura Polimeno*[1†], *Annalisa Coriolano*[1†], *Rosanna Mastria*[1*], *Francesco Todisco*[1],
*Milena De Giorgi*[1], *Antonio Fieramosca*[1], *Marco Pugliese*[1],
*Carmela T. Prontera*[1], *Aurora Rizzo*[1], *Luisa De Marco*[1], *Dario Ballarini*[1],
*Giuseppe Gigli*[1,2], *Daniele Sanvitto*[1*]

[1]CNR Nanotec, Institute of Nanotechnology, via Monteroni, 73100, Lecce, Italy.
[2]Dipartimento di Matematica e Fisica "Ennio de Giorgi", Universitá del Salento

# I  Fabrication of CsPbBr$_3$ single crystals microplatelets

<u>Synthesis of CsPbBr$_3$ microplatelets by capillary bridge approach.</u> To obtain CsPbBr$_3$ microplatelets, we use a silicon template consisting of structured square pillars (ThunderNil) (see Figure S1). To facilitate the confinement of the precursor solution on the top of the square pillars, we selectively modify the surface using heptadecafluorodecyltrimethoxysilane (FAS). This modification results in asymmetric wettability, featuring a hydrophobic sidewall and a lyophilic top wall. Initially, a thin layer of SU8 resist is spin-coated onto the surface of a flat silicon substrate. Subsequently, this substrate is brought into contact with the patterned silicon master. Gentle pressure is applied to the top, and placed at 70°C for 3 minutes on a hotplate. Next, the silicon master is transferred to a desiccator for a 24-hour period with 50 µl of FAS molecules and subsequently placed on a hot plate at 90°C for 2 hours. Finally, the substrate is washed with acetone, in order to remove the thin layer of the resist, obtaining an asymmetrical functionalized silicon master. We then sandwich a few microliters of the precursor solution between the template and the target substrate, which is previously treated with oxygen plasma for ten minutes.

---

[†] These authors contributed equally: L. Polimeno, A. Coriolano
[*] e-mail: rosanna.mastria@nanotec.cnr.it, daniele.sanvitto@nanotec.cnr.it

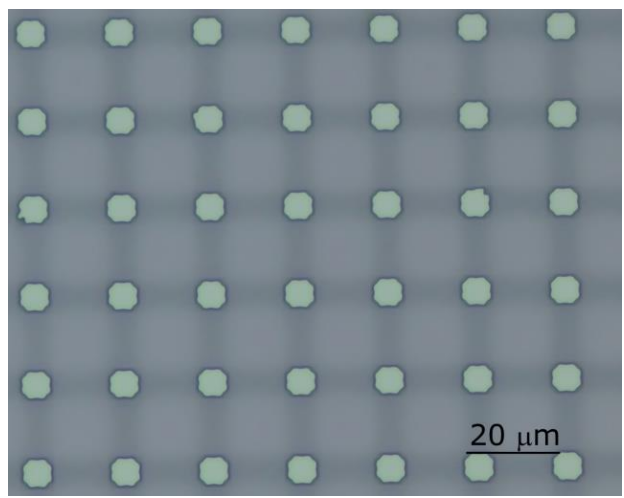

Figure S1: Optical image of the micro-structured silicon template.

A home-built apparatus is employed to apply uniform pressure across the sandwich, ensuring liquid confinement at the top of the pillars. The perovskite precursors ($PbBr_2$ and $CsBr$) in stoichiometric ratio are dissolved in Dimethyl sulfoxide (DMSO) at a concentration of 0.02 M under heating and stirring until a clear transparent solution was obtained. We explore various growth conditions to regulate the nucleation and crystallization of the $CsPbBr_3$ microplatelets (details in the following paragraph).

Investigating the Factors Influencing $CsPbBr_3$ Crystallization. Initially, we consider the effect of the antisolvent. Specifically, we exploit the antisolvent to achieve and control the supersaturation of the confined precursor solution. We investigate the effects of two different antisolvents: dichloromethane and chlorobenzene, by exposing the precursor solution to their respective vapors. Among the two antisolvents tested, we find that chlorobenzene is more effective in inducing controlled crystallization and promoting the formation of single crystals. In contrast, dichloromethane appears to promote the growth of larger, non-uniform crystals with defects at the edges (see Figure S2). This can be attributed to the different velocity at which these two antisolvents induce supersaturation in the precursor solution. Indeed, due to the lower boiling point of dichloromethane, the saturation of the growth environment with antisolvent vapors occurs more rapidly compared to chlorobenzene, resulting in accelerated and uncontrolled crystallization.

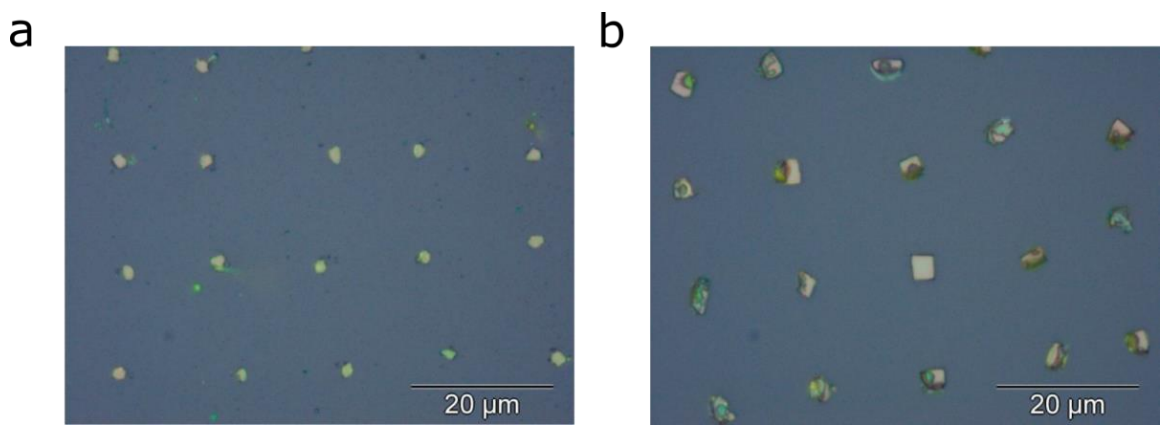

Figure S2: CsPbBr$_3$ microplatelets obtained by exposing the confined precursor liquid to antisolvent vapor: (a) Chlorobenzene;(b) Dichloromethane.

However, the vapor-based approach is not ideal. This often results in inhomogeneities in crystallization across the substrate. This is because vapors have difficulty penetrating effectively into the confined liquid, leading to inhomogeneities in crystallization across the substrate. To overcome this issue, we opt to include the antisolvent directly into the precursor solution. This allow for a better control over the supersaturation process while using a minimal quantity of the chosen antisolvent. Chlorobenzene (CB) is chosen as the optimal antisolvent because of its low dielectric constant ($\epsilon \sim 5.6$, respect to $\epsilon \sim 9$ of the dichlorometane) and the evident improved control over crystallization. A lower dielectric constant can favour the formation of precipitates because it reduces the solvent's ability to solvate the charged species present in the solution, thus favouring precipitation of the solute Initially, we then investigate the DMSO/CB (v/v) ratio to determine the saturation limit of CB in the 0.02 M solution (DMSO:CB=1:0.6).Subsequently, we assess the influence of DMSO/CB on the growth of $CsPbBr_3$ microplatelets. Specifically, we keep the ratio below the established limit to prevent instantaneous precipitation of the crystals. We find that the growth of the crystals is highly affected by variations in the solvent/antisolvent ratio. A DMSO:CB ratio of 1:0.3 resulted in the growth of uniformly-sized crystals, with some exhibiting a highly regular square shape. Conversely, increasing the ratio to 1:0.5 led to the precipitation of multicrystals with irregular shapes and position over the substrate (see Figure S3).

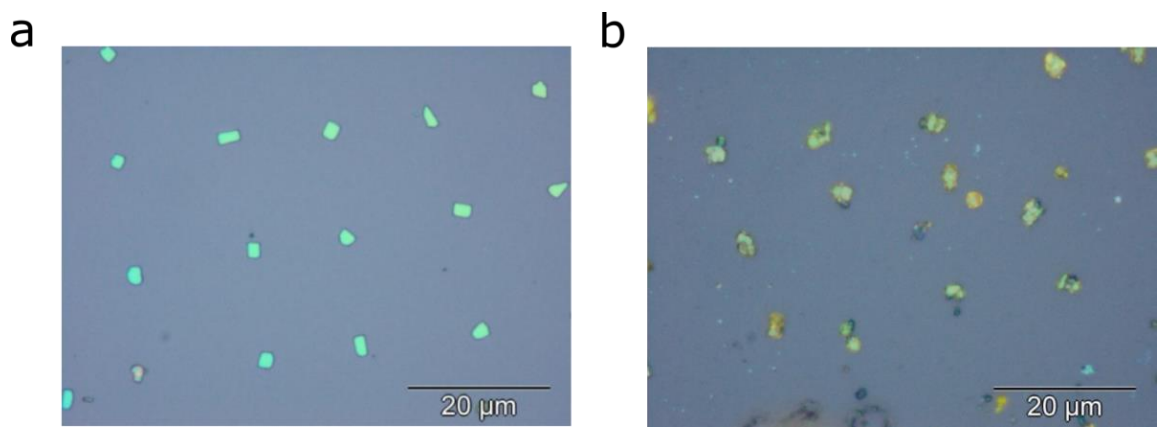

Figure S3: $CsPbBr_3$ microplatelets produced using varying DMSO/CB ratios: (a) DMSO:CB = 1:0.3; (b) DMSO:CB = 1:0.5.

Furthermore, we take into account the influence of temperature. Temperature is a key factor in determining the rate of solvent evaporation, which in turn influences the speed at which supersaturation and crystal formation occur. At high temperatures, such as 80°C, we observe the formation of multicrystalline structures with irregular thicknesses and shapes, likely due to rapid supersaturation and the emergence of multiple nuclei around each pillar. Conversely, lower temperatures hinder complete solvent evaporation due to the high boiling point of DMSO. To address this issue, we establish two distinct working temperatures: a lower one to promote the formation of nuclei (ideally one per pillar) and a higher one to promote the growth of these preformed nuclei. This two-temperature approach markedly enhances crystal quality and at simultaneously affects the dimensions of the final crystals. The lower the temperature, the smaller the crystal size (Figure S4).

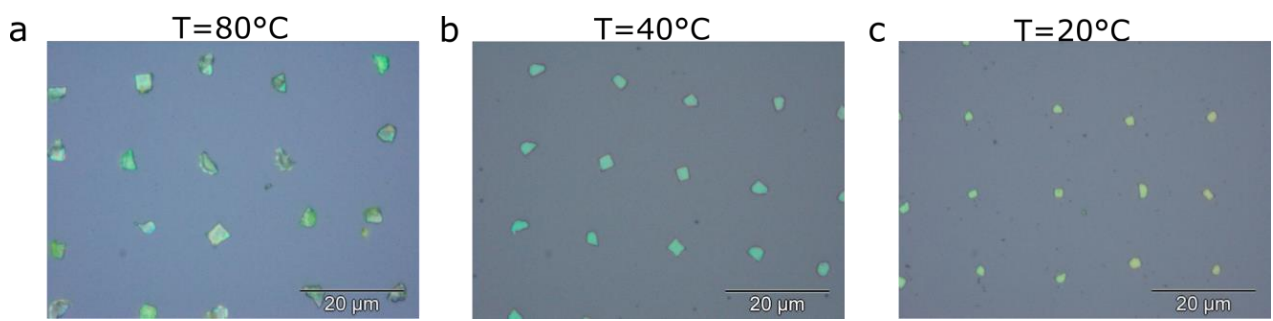

Figure S4: CsPbBr$_3$ microplatelets obtained by: (a) using a single high temperature of 80°C; (b) employing a two-temperature approach of 40°C for 48h followed by 80°C for 12h;(c) employing a two-temperature approach of 20°C for 48h followed by 80°C for 12h.

## II  Absorption Spectrum

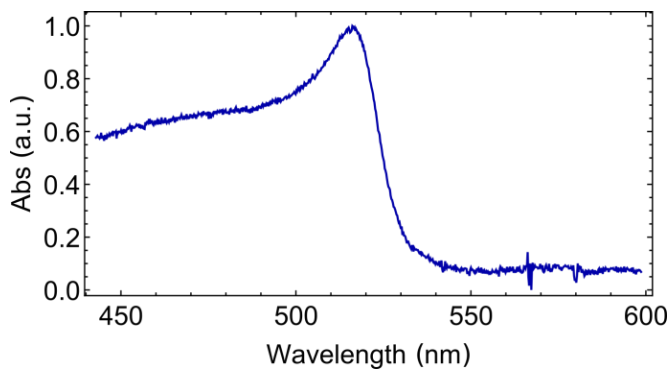

Figure S5: (a) Absorption spectrum of CsPbBr$_3$ single crystal. Tha absorption coefficient at the energy of the laser (2.64 eV - 470 nm) is 0.66.

# III  Whispering Gallery Modes

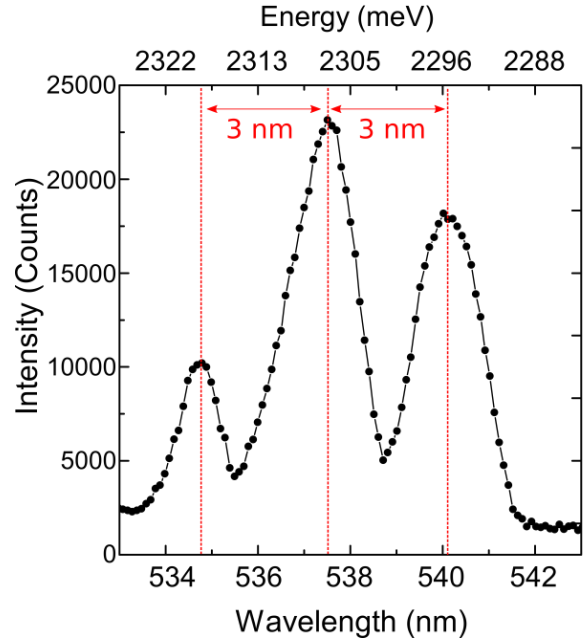

Figure S6: (a) Emission spectrum at higher fluence reported in the main text (F=330 µJ/cm²). The spacing between two adjacent mode is $\Delta\lambda \sim 3\ \mu m$.

To demonstrate the formation of WGM in CsPbBr$_3$ microplatelets, we extract the separation between two adjacent polariton modes from Fig. S6, yielding $\Delta\lambda \sim 3\ \mu m$. In a cubic WGM resonator, the mode spacing is determined by:

$$\Delta\lambda \cong \frac{\lambda^2}{2\sqrt{2}\cdot L \cdot n} \qquad (1)$$

where $n$ is the refractive index of the CsPbBr$_3$ single crystal ($n \sim 3.7$)[1], $L$ is the edge length of the microplatelts and $\lambda$ is the emission wavelength. The microplatelet's edge length, as presented in Fig. 1 of the main text, is $L \sim 6\ \mu m$. By applying equ. 1, the theoretical separation between two adjacent whispering gallery modes is calculated to be ~3.6 nm. This theoretical prediction aligns cohesively with the experimental value derived from the spectrum depicted in Fig. S6.

# IV  Fabrication of CsPbBr$_3$ single crystals microwires

<u>Synthesis of Cesium Nonanoate.</u> Cesium nonanoate is used as amphiphilic molecules to control the growth of CsPbBr$_3$ microwires as reported in previous report.[2]. To synthesize cesium nonanoate, 3.16 mmol of cesium carbonate (Cs$_2$CO$_3$) were mixed with 6.32 mmol of nonanoic acid (C$_9$H$_{18}$O$_2$) in 5 ml of ethanol solvent under a nitrogen flow. The temperature is then set and maintained at 100°C for about 30 minutes or until the mixture became completely clear. Subsequently, the ethanol is evaporated under vacuum at 100°C, and the resulting powder is kept under vacuum for 24 hours at room temperature to remove any residual solvent

<u>Synthesis of CsPbBr$_3$ microwires by capillary bridge approach.</u>

The CsPbBr$_3$ solution is prepared by dissolving CsBr and PbBr$_2$ in 1 ml of DMSO to obtain a 0.02 M solution. Subsequently, 20 µl of a 0.02 M solution of Cesium nonanoate in DMSO is added to the precursor solution. In this approach, the synthetic process is controlled using a periodic microwires structured silicon template. The sidewalls of the patterned silicon master (ThunderNil) are selectively modified with FAS molecules following a process similar to what was previously done for the growth of the perovskite microplatelets. Then, 1 µl of this solution is dispensed onto the desired substrate, which is previously treated with oxygen plasma for ten minutes. The target substrate is then covered with the asymmetrically functionalized silicon template. Solvent evaporation begins at 60°C in an oven with controlled humidity (RH 30%). After complete solvent evaporation, CsPbBr$_3$ microwire arrays are obtained on the target substrate (Figure S7).

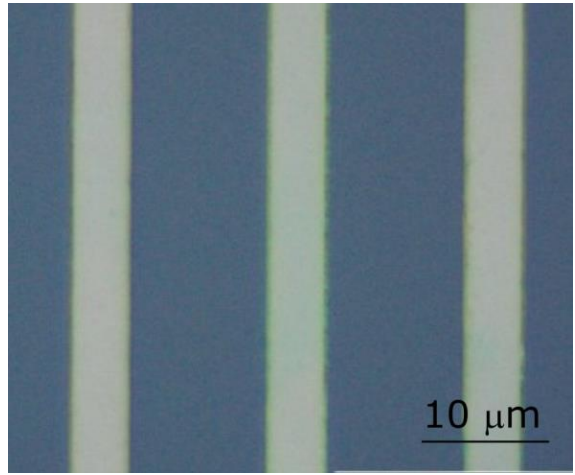

Figure S7: Optical image of CsPbBr$_3$ microwires on glass substrate.